\documentclass[12pt]{article}
 
\def\pa{\partial}

\def\a{\alpha}
\def\b{\beta}
\def\d{\delta} 
\def\e{\epsilon}

\def\l{\lambda} 
\def\m{\mu}
\def\n{\nu}

\def\o{\omega} 
\def\mn{{\mu\nu}}
\def\be{\begin{equation}}
\def\ee{\end{equation}}
\def\bm{\boldmath}

\begin{document}

\begin{center} {\Large\bf Clebsch (String) Parameterization\\[1ex] of
3-Vectors and Their Actions}
\bigskip

S. Deser\\[0.5ex]{\small\sl Department of Physics, Brandeis University,
Waltham, MA 02454, USA}
\medskip

R. Jackiw\\[0.5ex]{\small\sl Department of Physics, Massachusetts Institute
of Technology, Cambridge, MA 02139, USA}
\medskip
 
A.P. Polychronakos\\[0.5ex]{\small\sl Institute of Theoretical Physics, Uppsala
University, S-75108 Uppsala, Sweden\\
and Physics Department, University of Ioannina,
45110 Ioannina, Greece}\\[2ex]
MIT-CTP \# 2992 \quad---\quad Brandeis \# BRX TH-476
\end{center}

\begin{abstract}\noindent We discuss some properties of the intrinsically
nonlinear Clebsch decomposition of a vector field into three
scalars in $d=3$.  In particular, we note and account for the
incompleteness of this parameterization when attempting to use it
in variational principles  involving Maxwell and Chern-Simons
actions. Similarities with string decomposition of metrics and
their actions are also pointed out.
\end{abstract}

\section{Clebsch Decomposition}

The  decomposition of vectors, as well as higher rank tensors,
into irreducible parts is an ancient and extremely useful tool in
fluid mechanics, electrodynamics, and gravity. The longitudinal/transverse
split separates an arbitrary $d=3$ Euclidean vector into a scalar 
  plus a transverse vector,
\be
{\bf A} \equiv {\bf A}^L + {\bf A}^T = \mbox{\bm$\nabla$}\l + 
\mbox{\bm$\nabla$} \times {\bf W}^T  \; ,%
\ee%
the vector ${\bf W}^T$ being defined up to a gradient. This linear
orthogonal (upon spatial integration) parameterization naturally decomposes
the 3-vector {\bf A} into two, with 1 and 2 components, respectively. The
completeness of $(\mbox{\bm$\nabla$}\l,  {\bf W}^T)$ in representing {\bf
A} is evidenced by their uniqueness and invertibility (up to zero
modes); in particular, any variational principle yields the same
Euler-Lagrange system whether we vary {\bf A} or first decompose
it and then vary $(\l,  {\bf W}^T)$ separately.  It is easy to
check that since the field strength is
\be
 {\bf B} \mbox{\bm$\equiv$} \mbox{\bm$\nabla$} \times {\bf A} = -
\nabla^2 {\bf W}^T \; ,\ee%
the Chern-Simons (first used physically in \cite{woltier}) and
Maxwell actions
\be
I_{CS} ={\textstyle\frac{1}{2}}\: \m \: \int {\bf A}\cdot {\bf B}
= - {\textstyle\frac{1}{2}} \: \mu \: \int ( \nabla \times {\bf W}^T) \cdot (
\nabla^2 {\bf W}^T) \;\;\;\; I_M = -{\textstyle\frac{1}{2}} \:
\int {\bf B}^2 = - {\textstyle\frac{1}{2}}\: \int
(\nabla^2 {\bf W}^T)^2 \; ,
\ee%
vary with $A$~into
\be
\d I_{CS}/\d{\bf A} = \m {\bf B} \; , \;\;\;\; \d I_M/\d{\bf A} =
\mbox{\bm$\nabla$} \times {\bf B}%
\ee%
which agrees completely with the variations of (3) with respect to
$(\l, \;  {\bf W}^T)$.  The extra curl that appears upon varying
${\bf W}^T$ does not lose information: the curl is not a projector in the space
of transverse functions.

Another ancient parameterization, due to Clebsch \cite{clebsch},
has long been used in fluid mechanics \cite{berger}, but is less
familiar elsewhere.  This decomposition is strikingly different -- 
it involves only scalars rather than vectors and is nonlinear:
\be
{\bf A} = \mbox{\bm$\nabla$}\omega + \e^{ab} \phi_a
\mbox{\bm$\nabla$} \phi_b =\mbox{\bm $\nabla$}\omega^\prime +
2\phi_1\mbox{\bm $\nabla$} \phi_2 \; , \;\;\; {\bf B} =
\mbox{\bm$\nabla$} \times {\bf A} = \e^{ab} \mbox{\bm$\nabla$}
\phi_a \times
\mbox{\bm$\nabla$} \phi_b \; .%
\ee%
The representations (1) and (5) share the pure gauge longitudinal
term, but the role of   ${\bf W}^T$ has been
assumed by the pair of scalars $\phi_a$, $a$=1,2; $\e^{ab}=
-\e^{ba}$ is the alternating symbol.  The construction of these
scalars is described in the texts \cite{lamb}; for the moment we
note only that the $\phi_a$ are not uniquely determined by {\bf A}
or {\bf B}, nor can all ({\bf A,B}) be put in this form  in terms of well-defined
scalar functions. 

Our main interest is the particularly strange
phenomenon (not hitherto noted to our knowledge), that the Clebsch
decomposition (5) is incomplete in the very concrete sense that it fails
(differently for each), when inserted into the two actions of (3)
and varied with respect to $(\o , \phi_a)$, to give the field
equations (4) implied by the generic {\bf A}-variations. The
formal source of this incompleteness may be traced by comparing
the decompositions (1) and (5),
\be
\nabla^2\l = \mbox{\bm$\nabla$} \cdot {\bf A} = \nabla^2 \o +
\e^{ab} \phi_a \nabla^2 \phi_b \;, \;\;\;\;\;\; %
- \nabla^2 {\bf W}^T = \mbox{\bm$\nabla$} \times {\bf A} = \e^{ab}
\mbox{\bm$\nabla$} \phi_a \times
\mbox{\bm$\nabla$} \phi_b \; . %
\ee%
Note the demand of (6) that the curl of {\bf A} be expressible as
the cross-product of two gradients: the curl must then lie on the
line orthogonal to the plane they define. It follows that the
Chern-Simons action becomes a pure surface term,
\be
\int {\bf A} \cdot \mbox{\bm$\nabla$} \times {\bf A} = \int
\mbox{\bm$\nabla$} \o \cdot \mbox{\bm$\nabla$} \times {\bf A} =
\int \mbox{\bm$\nabla$} \cdot (\o \mbox{\bm$\nabla$} \times {\bf
A}) \; . %
\ee%
In $\o = 0$ gauge, {\bf A} is locally orthogonal to {\bf B}, and
in any gauge the bulk integral of (7) vanishes\footnote{When the volume 
integral of ${\bf A}\cdot\mbox{\bm$\nabla$} \times{\bf A}$ is nonzero, its
entire contribution in the Clebsch parameterization for~$\bf A$ must come
from surface terms. Consequently, the Clebsch potentials $\omega, \phi_a$
must be ill behaved, either in the finite volume or at infinity, so that the
surface integrals do not vanish. For an example and further details,
see~\cite{pi}}. {\it A fortiori}, the variation of $I_{CS}$ with respect to $(\o ,
\phi_a)$ is trivial.  Pure Chern-Simons dynamics here reduces entirely to the
boundary\footnote{In this respect, when CS is used as the ``helicity" in
magnetohydrodynamics, it bears a resemblance to gravitational energy; there
is no sensible local (gauge invariant) density, but a perfectly well defined
integral over the boundary.}, rather than defining the (trivially ${\bf B}$=0)
behavior in the bulk.

Variation of the Maxwell action is also incomplete:
\be
-\d \textstyle{\frac{1}{2}} \int {\bf B}^2 \equiv -\int {\bf B}
\cdot \e^{cd} (\mbox{\bm$\nabla$} \phi_c \times \mbox{\bm$\nabla$}
\d \phi_d ) \; , %
\ee%
resulting in the pair of Euler-Lagrange equations
\be\label{eq:9}
(\mbox{\bm$\nabla$} \times {\bf B}) \cdot \mbox{\bm$\nabla$}
\phi_a \equiv \mbox{\bm$\nabla$} \cdot (\phi_a \mbox{\bm$\nabla$}
\times {\bf B})
= 0\; , \;\;\;\;\;\;\;\; a = 1,2 \; .\ee%
That is, $\mbox{\bm$\nabla$}\times {\bf B}$, like {\bf B} itself,
is orthogonal to the plane defined by $(\mbox{\bm$\nabla$}
\phi_a)$: the field equation is then
\be\label{eq:10}
\mbox{\bm$\nabla$} \times {\bf B} + \m ({\bf r}){\bf B} = 0 \;. %
\ee%
The function of integration, $\m ({\bf r})$, is constrained by the
transversality of $({\bf B}, \mbox{\bm$\nabla$}\times {\bf B})$ in
(10) to have its gradient orthogonal to them both:
$\mbox{\bm$\nabla$} \mu$ must lie in the ($\mbox{\bm$\nabla$}
\phi_a$) plane.

The indeterminacies in the two variational principles are rather
different: from $I_{CS}$, there is no field equation at all; this
is in fact traceable to the Darboux incompleteness of the Clebsch
decomposition \cite{sternberg}, which rests on the triviality of the bulk
contribution to 
$I_{CS}$ that it imposes$^{1}$. From the Maxwell
action, we obtain only the direction of $\mbox{\bm$\nabla$}\times{\bf B}$.
Indeed, any gauge invariant action $I[{\bf B}]$ will provide the same form
\be\label{eq:11}
[\mbox{\bm$\nabla$}\times \d I/\d {\bf B}(x)] \cdot
\mbox{\bm$\nabla$} \phi_a
= 0 \; ,%
\ee%
and hence allow the same indeterminate $\m (r) {\bf B}$ addition,
subject to $\mbox{\bm$\nabla$} \m \cdot {\bf B}=0$.

The topologically massive-like \cite{templeton} ``Maxwell"
equation (10) is quite unusual:  Consider (9) in ``covariant"
notation,%
\be\label{eq:12}
(\pa_\n F^\mn ) \pa_\m \phi^a \equiv
\pa_\m (\pa_\n F^\mn \phi^a ) = 0 %
\ee%
whose solution is %
\be\label{eq:13}
\pa_\n F^\mn \phi^a = \e^{\nu\a\b} \pa_\a Z^a_\b %
\ee%
where $Z^a_\a$ are two arbitrary transverse 3-vectors. This means in
turn that there are two (conserved) ``currents'', for example,
$j^\m=(\phi_b\phi^b)^{-1} \phi_a \e^{\m\a\b}\pa_\a Z^a_\b.$
Equations~(\ref{eq:9},\ref{eq:10}) and ~(\ref{eq:12},\ref{eq:13}) present the
field equations in two equivalent ways. If we instead introduce the explicit
form (6) into (9), we get two (cubic) equations in $\phi$ alone,%
\be
\pa_\n [ \e^{bc} \phi^b (\pa^2_\mn - \d_\mn \nabla^2 )\phi^c ]
\pa_\m \phi^a = 0 \; .  %
\ee%


A nonabelian generalization of the Clebsch decomposition has
recently been suggested \cite{jackiw}.  It shares the abelian
property that the CS term is a pure divergence and a similar
degeneracy in the field equations occurs.  The basic mechanism in
both cases is that the variation of an action with respect to the
initial variable, here $A_\m$ or its color generalization, is to
be multiplied by $\d A_\m / \d\o , \; \d A_\m /\d \phi^a$. The
degenerate contributions noted above again occur as a result of
the $\phi^a$ variations.

It is amusing to compare the Clebsch vector decomposition to the
``string-embedding" partition of a 2-tensor \cite{regge},
available in any dimension $d$,
\be%
 g_\mn (x) = y^A_\m (x) y^B_\n (x) \eta_{AB} \; , \;\;\; %
 y^A_\m \equiv \pa_\m y^A \; .  %
\ee%
The scalar string variables $y^A$, with the index $A$ ranging over
$1,\dots,\frac{1}{2} d(d+1)$, are now the independent ones.
Inserting this decomposition into the Einstein action (or indeed
any other covariant one) and varying the $y^A$,%
$$%
\d I = \int G^\mn \d g_\mn / \d y_A \d y_A \equiv - \int D_\m D_\n
(G^\mn y^A) \d y^A \eqno{(16\rm{a})}
$$%
(where $G^\mn \equiv \d I/\d g_\mn$ is the necessarily identically
conserved ``Einstein" tensor density of the action $I[g]$) gives
Euler-Lagrange equations very similar to the ``weak Maxwell"
ones, in terms of the covariant derivative $D_\m (g)$
$$%
G^\mn (g(y))  D_\m D_\n y^A = 0 \; . \eqno{(16\rm{b})}
$$%
These are much weaker \cite{regge} than\footnote{One might
conjecture that a quantity similar to the Chern-Simons form
becomes degenerate in the metric case, for example, that yielding
the Cotton tensor (see \cite{templeton}) in $d=3$.  The absence of
CS generalizations in even~$d$ may account for the lack of a higher
$d$ Clebsch representation for vectors.  For odd $d$, the
extension of (5), $A=\mbox{\boldmath$\nabla$}\o + \sum^{d-2}_{i=1}
\phi_i\mbox{\boldmath$\nabla$} \phi_{i+1}$, trivializes the
corresponding CS $\sim A\cdot F_1\cdots F_{d-2}$.} $G^\mn = 0$. In
both vector and tensor cases, the problem is with the derivative
nature of a parameterization by scalars: the equally nonlinear
\emph{vielbein} decomposition of the metric $g_\mn = e^a_\m e^b_\n
\eta_{ab}$ into the set of vectors $(e_\m )^a$ is of course perfectly
acceptable (if the \emph{vielbeins} are invertible). Also true of both
the Clebsch and the tensor decompositions is the inaccessibility
of weak field, linearized, theory from these nonlinear
parameterizations, unlike the (invertible) \emph{vielbein} choice, where
there is a vacuum possibility $e^a_\m = \d^a_\m$ about which to
expand.

This work was supported in part by National Science Foundation
grant PHY99-73935 and US~Department of Energy (DOE) contract
\#~DE-FC02-94ER40818.

 \end{document}